# Luminescence thermometry based on time gates: highly sensitive approach for real time sensing and imaging

M. Szymczak[1], D. Szymański[1], M. Piasecki[2], M. Brik[2,3,4,5,6], L. Marciniak[1*]

[1] Institute of Low Temperature and Structure Research, Polish Academy of Sciences, Okólna 2, 50-422 Wrocław, Poland

[2] Faculty of Science and Technology, Jan Długosz University, Armii Krajowej 13/15, 42-200 Częstochowa, Poland

[3] School of Optoelectronic Engineering, Chongqing University of Posts and Telecommunications, Chongqing 400065, China

[4] Centre of Excellence for Photoconversion, Vinča Institute of Nuclear Sciences - National Institute of the Republic of Serbia, University of Belgrade, Belgrade, Serbia

[5] Institute of Physics, University of Tartu, W. Ostwald Str. 1, 50411 Tartu, Estonia

[6] Academy of Romanian Scientists, Ilfov Str. No. 3, 050044 Bucharest, Romania

*corresponding author: l.marciniak@intibs.pl

KEYWORDS: luminescence thermometry, optical sensors, lifetime-based approach, thermal imaging, temporal rates

## Abstract

Undoubtedly, one of the most significant advantages of luminescence thermometry is its ability to be used not only for spot temperature measurements but also for imaging temperature changes. Among the commonly proposed approaches, luminescence thermometry based on luminescence kinetics holds particular promise. However, most thermometric studies rely on the analysis of luminescence decay profiles, a method that significantly hinders, if not entirely precludes, real-time thermal imaging. In this paper, we propose an alternative approach based on the luminescence intensity ratio integrated over two temporal gates. Tests conducted on two representative phosphors, $Ba_2LaNbO_6$:1%$Mn^{4+}$ and $Ca_2LaNbO_6$:1%$Mn^{4+}$, demonstrate that the proposed method not only enables thermal imaging but also achieves substantially higher relative sensitivity, reaching $S_R$=17.1 % $K^{-1}$ for $Ba_2LaNbO_6$:1%$Mn^{4+}$ and $S_R$=9.4 % $K^{-1}$ for $Ca_2LaNbO_6$:1%$Mn^{4+}$, compared to the conventional lifetime-based approach ($S_R$=4.2 % $K^{-1}$ for $Ba_2LaNbO_6$:1%$Mn^{4+}$ and $S_R$=1.2 % $K^{-1}$ for $Ca_2LaNbO_6$:1%$Mn^{4+}$). Furthermore, careful selection of gate lengths allows optimization of the thermometric performance of the proposed luminescent thermometers. This approach enables expansion of the thermal operating range at the cost of relative sensitivity, providing versatility to adapt the thermometer for specific applications.

**Introduction**

While numerous spectroscopic parameters of phosphor exhibit sensitivity to temperature changes and can therefore be utilized in luminescence thermometry, ratiometric and lifetime-based approaches are undoubtedly the most widely employed[1–10]This is due to several key advantages. Most notably, these methods provide highly reliable temperature readings, exhibit insensitivity to other experimental parameters (unlike intensity-based approaches), and do not require high-resolution spectrometers (as needed in spectral shift- or bandwidth-based approaches)[2,11]. Furthermore, both techniques enable not only point-based thermal readouts but also imaging of temperature distributions[12–18]. The effectiveness of the ratiometric approach has been well demonstrated, both using bandpass optical filter sets[16,19] and in camera-only approaches[13] leveraging the relationship between images captured by the RGB channels of a standard sensor. This capability is a significant advantage, as it allows the use of low-cost cameras, including smartphones, for temperature readout. However, as many studies have shown, the dispersive nature of the extinction coefficient of the medium in which the phosphor is located-or that exists in the optical path between the detector and the luminescence thermometer-can distort the phosphor's emission spectrum, potentially compromising the reliability of the readings[11]. Dynamic factors such as humidity or gas content in the optical path can further complicate this issue, making in-situ thermometer recalibration ineffective in many cases.

For these reasons, a lifetime-based approach may offer a better solution[2,6,11]. In this approach, detection occurs at a single emission wavelength, and the thermometer parameter is derived from the temporal evolution of the luminescence signal. As consistently demonstrated, luminescence kinetics is highly effective for temperature sensing applications[20–22] and recent reports have demonstrated that acquisition times as short as tens of milliseconds are sufficient to obtain a lifetime-based thermal readout[23]. However, the sensitivity of the lifetime-based method is typically lower than that of its ratiometric counterpart[3,6] Moreover, while lifetime-based thermometry is popular in the literature, it is predominantly reliant on luminescence decay profile measurements[24–32]. This approach, while effective, precludes real-time thermal imaging, as it necessitates point-by-point measurements of the luminescence decay profile, which is exceedingly time-consuming.

To address these limitations, this study explores a ratiometric approach to luminescence kinetics based on time-gated temperature determination. Given the rare application of this method in luminescence thermometry, Figure 1 schematically illustrates its potential practical implementation. This study explores the capabilities of this approach without presenting a proof-of-concept experiment. Nonetheless, its efficacy in temperature measurement has been previously validated[33]. Two phosphors doped with $Mn^{4+}$ ions were used as representative examples: $Ba_2LaNbO_6$:1%$Mn^{4+}$ and $Ca_2LaNbO_6$:1%$Mn^{4+}$. Although the proposed approach is universal and can be used for any type of phosphor, there are several important advantages of materials doped with $Mn^{4+}$ ions that facilitate the demonstration of the advantages of this solution[10,34–45]. A notable advantage of $Mn^{4+}$ ions in this context is that their $^2E \rightarrow {^4A_2}$ electron transition is spin-forbidden, resulting in a long luminescence decay profile. Additionally, the luminescence of $Mn^{4+}$ ions is characterized by high intensity. Moreover, the change in the strength of the crystal field interacting with $Mn^{4+}$ ions enables modification of the rate of thermal shortening of the $^2E$ state[36–43]. In this approach thermal sensing or/and imaging of the analysed object covered with luminescence thermometer can be determined by the analysis of the luminescence intensity integrated into two temporal gates. The use of spatial detector like digital camera enables the two luminescence maps to be captured. Their luminescence intensity ratio can be easily transformed into thermal maps using the calibration curve. Comparative analysis of relative sensitivities using $\tau_{avr}$-based methods and the intensity ratio approach recorded in time gates demonstrates the superiority of the latter. Furthermore, this study highlights how the thermometric performance of such a thermometer can be modulated and optimized by adjusting the timing of the gates and the delay between them. The approach of utilizing the ratio of luminescence intensities recorded at two distinct time gates was already extensively employed in imaging various physical and chemical parameters[46–50]. However, its application in luminescence thermometry remains relatively uncommon. Previous studies have explored this method with phosphors doped with $Cr^{3+}$ [51,52] and lanthanide ions [52,53]. The high relative sensitivity values achieved with $Mn^{4+}$-doped phosphors, as presented in this work, render this approach particularly promising for practical applications.

## 2. Experimental Section

*Synthesis*

Phosphors $Ba_2LaNbO_6$ and $Ca_2LaNbO_6$ doped with 1% of $Mn^{4+}$ ions were synthesized using a conventional, high-temperature solid-state reaction method. The dopant ion concentration was selected arbitrarily based on our previous experience to balance high emission intensity and absorption cross-section while minimizing the risk of structural defects arising from ionic charge mismatches between the dopant and host cations.

Stoichiometric amounts of $Ba(NO_3)_2$ or $CaCO_3$ (both with 99.999% purity, Thermo Scientific Chemicals), $La_2O_3$ (99.99% of purity, Stanford Materials Corporation), $Nb_2O_5$ (99.9985% of purity, Thermo Scientific Chemicals) and $MnCl_2·4H_2O$ (>99.0% of purity, Sigma Aldrich) were carefully weighted, with the amount of the $MnCl_2·4H_2O$ calculated relative to the $Nb^{5+}$ ions in the host material. The powders were thoroughly ground in an agate mortar, using a 5 ml of hexane (3 times) to ensure better homogeneity of the mixture. The resulting powders were then transferred to alumina crucibles and calcined in air at 1573 K for 6 hours, with heating rate of 10 K min$^{-1}$. Afterward, the obtained phosphors were cooled naturally to room temperature and ground once again for subsequent structural and spectroscopic studies.

*Methods*

The X-ray diffraction (XRD) analysis of obtained powders was performed using a PANalytical X'Pert Pro diffractometer equipped with an Anton Paar TCU1000 N Temperature Control Unit using Ni-filtered Cu Kα radiation (V = 40 kV, I = 30 mA). Measurement were performed in the 2theta=10-70º range with the 0.02626º step and 60 min measurement time.

The Scanning Electron Microscopy (SEM) was used to verify the morphology of the samples and the distribution of its elements by EDS mapping. The FEI Nova NanoSEM 230 equipped with an EDAX Genesis XM4 energy dispersive spectrometer was used for measurements (V=30 kV for SEM and V=5 kV for EDS mapping). Samples were prepared by dispersing a some amount of powder in a few drops of methanol. A drop of the resulting suspension was placed on the carbon stub and dried.

The spectroscopic analysis (emission/excitation spectra and luminescence kinetics) was performed with the use of FLS1000 Fluorescence Spectrometer form Edinburgh Instruments, equipped with the 450 W Xenon lamp and R5509-72 photomultiplier tube from Hamamatsu with nitrogen-flow cooled housing as the detector. The temperature-dependent measurements were carried our using the THMS 600 heating-cooling stage from Linkam, which provides a temperature stability of 0.1 K and a set point resolution of 0.1 K. Before each measurement, the temperature was stabilized for 2 minutes to ensure reliable readouts.

The average luminescence lifetime ($\tau_{avr}$) was calculated using Equation 1, based on double-exponential fit of the luminescence decay curves (Eq. 2):

$$\tau_{avr} = \frac{A_1\tau_1^2 + A_2\tau_2^2}{A_1\tau_1 + A_2\tau_2} \qquad (1)$$

$$I(t) = I_0 + A_1 \cdot \exp\left(-\frac{t}{\tau_1}\right) + A_2 \cdot \exp\left(-\frac{t}{\tau_2}\right) \qquad (2)$$

To determine the $S_R$ the thermal dependencies of $\tau_{avr}$ and LIR were fitted using dose-response curve (eq. 1 in SI). The experimental values of $S_R$ were calculated using the obtained fitting curves at each $T_i$ temperature as follows:

$$S_R(T_i) = \frac{100\%}{\Omega}\frac{\Omega(T_i) - \Omega(T_{i-1})}{T_i - T_{i-1}} \qquad (3)$$

where $\Omega(T_i)$ represents the $\tau_{avr}$ or LIR at temperature $T_i$.

**Results and discussion**

Both $Ba_2LaNbO_6$ and $Ca_2LaNbO_6$ belong to the class of double perovskite compounds with the general formula $A_2BB'O_6$, where $A$ usually represents elements from the $s$ or $p$-blocks, while $B$ and $B'$ are elements from the $d$ or $f$-blocks[54–66]. These compounds crystalize in a monoclinic crystal structure but differ in their space groups: $I2/m$ [54] for $Ba_2LaNbO_6$ and $P2_1/c$

[67] for $Ca_2LaNbO_6$. In general, both structures consist of $Ca^{2+}$ or $Ba^{2+}$ and $La^{3+}$ ions in 6- or 7-fold coordination of $O^{2-}$ anions (Figure 1a and 1b). In $Ba_2LaNbO_6$, all cations are octahedrally coordinated, whereas in $Ca_2LaNbO_6$, $Ca^{2+}$ and $La^{3+}$ form both 6-fold and 7-fold coordinated polyhedral[54–66,68–71]. In both structures, $Nb^{5+}$ ions form $(NbO_6)^{7-}$ octahedra. XRD analysis confirmed that the synthesized samples contain a single expected phase, with no additional reflections indicating impurities, as compared to the ICSD No. 172403 and JCPDS No. 70-1157 reference patterns for $Ba_2LaNbO_6$ and $Ca_2LaNbO_6$, respectively (Figure 1c, see also Figure S1). For $Ca_2LaNbO_6$, the obtained reflections were compared to the analogous $Ca_2LaTaO_6$ monoclinic structure due to the unavailability of a suitable reference patterns in crystallographic database. The matching reflections with the reference data also indicate the successful incorporation of $Mn^{4+}$ ions into the $Ba_2LaNbO_6$ and $Ca_2LaNbO_6$ structures at molar amount of 1%. The $Mn^{4+}$ ions preferentially occupy crystallographic positions that provide six-fold coordination with ligands[39,72,73]. In both $Ba_2LaNbO_6$ and $Ca_2LaNbO_6$ structures, all cations can form octahedra. Consequently, determining the most probably crystallographic site for $Mn^{4+}$ incorporation requires consideration of the ionic charge and difference between ionic radius of the host and dopant cations. Regarding ionic charge, none of the cations in studied structures are characterized by 4+ charge. $Ba^{2+}$ and $Ca^{2+}$ can be excluded as potential sites for $Mn^{4+}$ due to the significant charge mismatch, which would make such substitution energetically unfavorable. This mismatch would result in charge compensation, potentially causing structural defects or oxidation of cations to higher oxidation states. Although $Nb^{5+}$ and $La^{3+}$ ions also differ in ionic charge from $Mn^{4+}$, the difference is smaller compared to other cations in the host structures. Consequently, the probability of $Mn^{4+}$ substitution in these positions is significantly higher. In a case of the difference in ionic radii ($D_R$) between the host ion ($R_H$) and dopant ion ($R_D$), it can be calculated using the following equation:

$$D_R = \frac{(R_H - R_D)}{R_H} \cdot 100\% \qquad (4)$$

For successful incorporation of a dopant ion into host lattice, the difference in ionic radii must be less than 30%. In this case, this condition is fulfilled only for the $Nb^{5+}$ ions, with a difference of 17.19% compared to $Mn^{4+}$ ions, as summarized in Table 1. Consequently, in both $Ba_2LaNbO_6$ and $Ca_2LaNbO_6$ structures, $Mn^{4+}$ ions are most probably located at $Nb^{5+}$ crystallographic positions, which provide the most favorable crystallographic environment. This hypothesis is also supported by the observed shift in the XRD reflections of the studied phosphors towards larger angles compared to their reference patterns. This shift indicates a reduction in the crystallographic unit cell size, associated with the substitution of smaller $Mn^{4+}$ ions (0.53 Å) for $Nb^{5+}$ ions, which have a larger ionic radius of 0.64 Å [74].

**Table 1.** Ionic radii of all 6-coordinated host cations and the $Mn^{4+}$ dopant ions, along with the calculated differences between them.

| Cation (6-coordinated) | Ionic radius (Å) | $D_R$ (%) |
|---|---|---|
| $Ba^{2+}$ | 1.35 | 60.74 |
| $Ca^{2+}$ | 1.00 | 47.00 |
| $La^{3+}$ | 1.03 | 48.64 |
| $Nb^{5+}$ | 0.64 | 17.19 |
| $Mn^{4+}$ | 0.53 | - |

SEM imaging revealed the presence of predominantly macrocrystals in both $Ba_2LaNbO_6$ and $Ca_2LaNbO_6$:1%$Mn^{4+}$ powders (Figure 1d and 1e, respectively). The crystals lack a uniform, characteristic shape and tend to aggregate, but their average size was estimated to be 3.0 ± 0.2 µm. Obtained morphology is a result of the high-temperature solid-state synthesis method used.

One of the key requirement for reliable spectroscopic characterization is the homogeneity of the samples in terms of elemental composition, especially dopant ions. Therefore, this feature was investigated using EDS analysis, which resulted in elemental distribution maps for Ba or Ca, La, Nb, O, and Mn. As shown in Figure 1d and 1e, both samples exhibit a high degree of elemental homogeneity, with an even distribution of all elements throughout the entire crystal studied. Furthermore, the EDS analysis confirmed the successful incorporation of $Mn^{4+}$ dopant ions into the host structures.

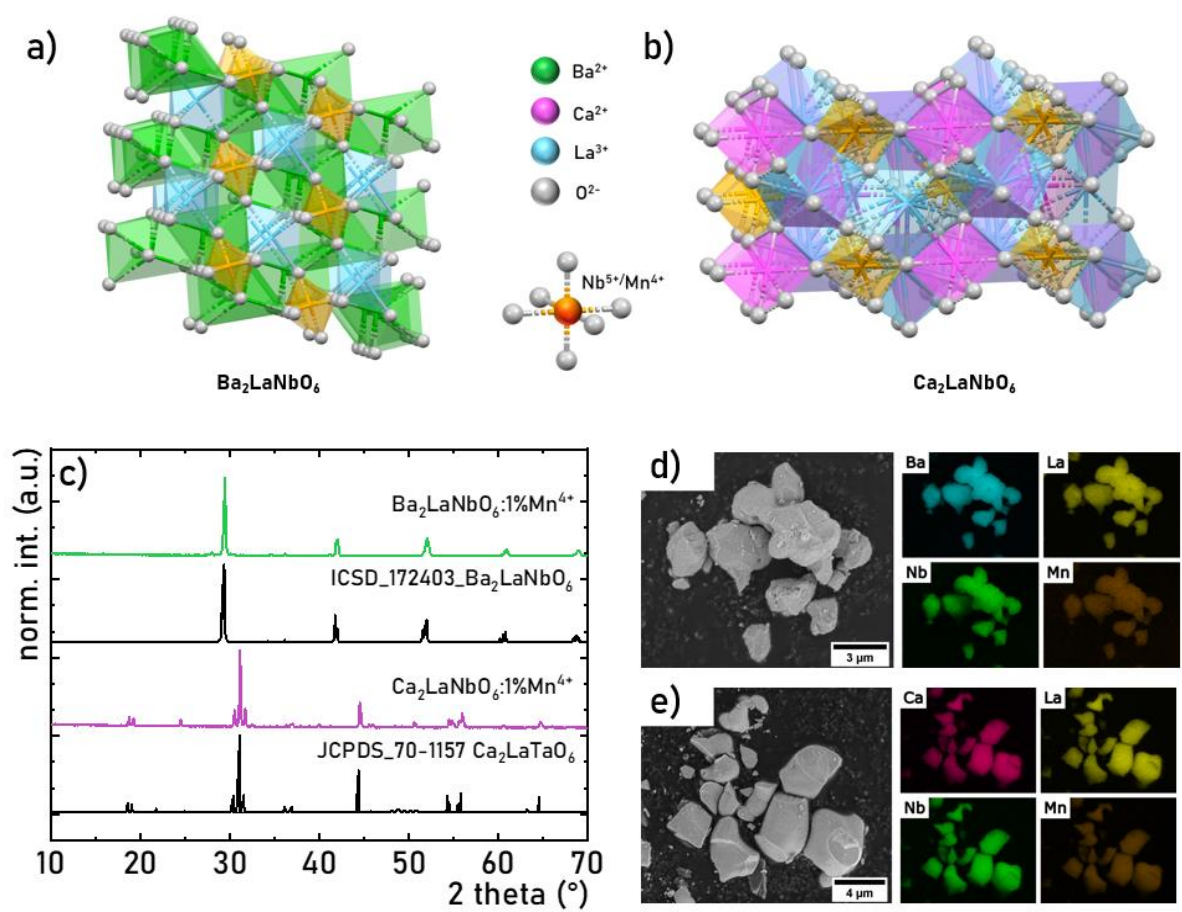

**Figure 1.** Visualization of $Ba_2LaNbO_6$ -a) and $Ca_2LaNbO_6$ -b) structures; comparison of the XRD patterns of $Ca_2LaNbO_6$:1%$Mn^{4+}$ and $Ca_2LaNbO_6$:1%$Mn^{4+}$ powders -c); representative SEM image and corresponding elemental maps of the $Ba_2LaNbO_6$:1%$Mn^{4+}$-d) and $Ca_2LaNbO_6$:1%$Mn^{4+}$-e).

The luminescence of $Mn^{4+}$ ions arises from the depopulation of the $^2E$ level to the $^4A_2$ ground state[34,36–43]. The $^2E \rightarrow {}^4A_2$ electron transition is spin-forbidden and corresponds to photon emission in the red spectral range (Figure 2a). According to the Tanabe-Sugano diagram for transition metal ions with a $3d^3$ electron configuration, the energy of the $^2E$ level is nearly independent of the crystal field strength[75,76]. Therefore, variations in the crystal field strength acting on $Mn^{4+}$ ions due to changes in the host material composition have minimal impact on the spectral position of the $^2E \rightarrow {}^4A_2$ emission band. However, the energy of the $^2E$ level depends on the covalency of the $Mn^{4+}$ - $O^{2-}$ bond, leading to slight shifts in the spectral position of $Mn^{4+}$ emission with changes in the host material. While the spectral position of the $^2E \rightarrow {}^4A_2$ band is not highly sensitive to changes in crystal field strength, variations in this parameter can influence the thermal stability of $Mn^{4+}$ luminescence. The primary thermally induced depopulation mechanism of the $^2E$ level involves crossover processes between the $^2E$ and $^4T_2$ state parabolas. Since the energy of the $^4T_2$ level increases with stronger $Dq/B$ crystal fields- and $Dq/B$ is proportional to $R^{-5}$ (where $R$ is the $Mn^{4+}$-$O^{2-}$ distance)-materials with shorter $Mn^{4+}$-$O^{2-}$ distances are expected to exhibit higher thermal stability of luminescence[42]. The energy of the $^4T_2$ level also affects the luminescence kinetics of the $^2E$ level. Spin-orbit coupling between the $^2E$ and $^4T_2$ levels partially relaxes the $^2E \rightarrow {}^4A_2$ electron transition, as the $^4T_2 \rightarrow {}^4A_2$ transition is spin-allowed. Consequently, materials with stronger crystal fields exhibit a larger energy separation between the $^2E$ and $^4T_2$ levels, reducing spin-orbit coupling and prolonging the lifetime of the $^2E$ state. A comparison of the luminescence spectra of $Ba_2LaNbO_6$:1%$Mn^{4+}$ and $Ca_2LaNbO_6$:1%$Mn^{4+}$ reveals notable differences in the spectroscopic properties of $Mn^{4+}$ ions. For $Ba_2LaNbO_6$:1%$Mn^{4+}$, at 83 K, the zero-phonon line and vibrational lines associated with the $^2E \rightarrow {}^4A_2$ transition are distinctly observable (Figure 2b, Figure S2). In contrast, for $Ca_2LaNbO_6$:1%$Mn^{4+}$, vibrational lines are not distinguishable in the spectrum, even at low temperatures and under low excitation power density, likely due to inhomogeneous broadening caused by localized distortions of the crystallographic sites occupied by $Mn^{4+}$ ions[64,68,69]. The higher energy of the $^2E \rightarrow {}^4A_2$ band for $Ba_2LaNbO_6$:1%$Mn^{4+}$ (15,120 cm$^{-1}$) compared to $Ca_2LaNbO_6$:1%$Mn^{4+}$ (14,200 cm$^{-1}$) suggests that the average $Mn^{4+}$-$O^{2-}$ distance is greater in $Ba_2LaNbO_6$:1%$Mn^{4+}$. A comparison of the excitation spectra of $Ba_2LaNbO_6$:1%$Mn^{4+}$ and $Ca_2LaNbO_6$:1%$Mn^{4+}$, measured at 83 K, indicates that both spectra consist of five bands associated with $^4A_2 \rightarrow {}^4T_2$ electronic transitions (20,0062 cm$^{-1}$ for $Ba_2LaNbO_6$:1%$Mn^{4+}$ and 19,783 cm$^{-1}$ for $Ca_2LaNbO_6$:1%$Mn^{4+}$), $^4A_2 \rightarrow {}^2T_2$ electronic transitions (26,921 cm$^{-1}$ for $Ba_2LaNbO_6$:1%$Mn^{4+}$ and 24,635 cm$^{-1}$ for $Ca_2LaNbO_6$:1%$Mn^{4+}$), $^4A_2 \rightarrow {}^4T_1$ electronic transitions (29,180 cm$^{-1}$ for $Ba_2LaNbO_6$:1%$Mn^{4+}$ and 27,170 cm$^{-1}$ for $Ca_2LaNbO_6$:1%$Mn^{4+}$), $Mn^{4+} \rightarrow O^{2-}$ electronic transitions (31,684 cm$^{-1}$ for $Ba_2LaNbO_6$:1%$Mn^{4+}$ and 29,182 cm$^{-1}$ for $Ca_2LaNbO_6$:1%$Mn^{4+}$) and a group of electronic transitions from the ground $^4A_2$ state to the closely located orbital triplets and doublets (all are split by low-symmetry crystal field) coming from the $^2G$, $^2H$, $^2P$ terms of the $Mn^{4+}$ ions (35,139 cm$^{-1}$ for $Ba_2LaNbO_6$:1%$Mn^{4+}$ and 32,170 cm$^{-1}$ for $Ca_2LaNbO_6$:1%$Mn^{4+}$) (Figure 2c, Figure S3 and S4). Higher values for the $^4A_2 \rightarrow {}^4T_2$ and $^4T_1$ band energies in $Ba_2LaNbO_6$:1%$Mn^{4+}$ suggest a stronger crystal field acting on $Mn^{4+}$ ions in this host material compared to $Ca_2LaNbO_6$:1%$Mn^{4+}$. To verify this hypothesis, the crystal field strength ($Dq/B$) and Racah parameters ($B$ and $C$) were determined using the following equations[39,77]:

$$E\left(^4A_2 \rightarrow {}^4T_2\right) = 10Dq \tag{5}$$

$$\frac{Dq}{B} = \frac{15\left(\frac{\Delta E}{Dq} - 8\right)}{\left(\frac{\Delta E}{Dq}\right)^2 - 10\frac{\Delta E}{Dq}} \tag{6}$$

$$\frac{E\left(^2E \rightarrow {}^4A_2\right)}{B} = \frac{3.05C}{B} + 7.9 - \frac{1.8B}{Dq} \tag{7}$$

The calculated values of $Dq/B$, $B$, and $C$ were 3.02, 662, 3370, for $Ba_2LaNbO_6$:1%$Mn^{4+}$ and 2.73, 723, 3068 for $Ca_2LaNbO_6$:1%$Mn^{4+}$ (Figure 2d). These results confirm the stronger crystal field interacting with $Mn^{4+}$ ions in $Ba_2LaNbO_6$:1%$Mn^{4+}$. Based on the Racah parameters, the covalency parameter was determined using the equation[78]:

$$\beta_1 = \sqrt{\left(\frac{B}{B_0}\right)^2 + \left(\frac{C}{C_0}\right)^2} \qquad (8)$$

where $B_0$ and $C_0$ are the values for free ions equal to 1160 and 4303 cm$^{-1}$, respectively. The lower $\beta_1$=0.9472 value for $Ca_2LaNbO_6$:1%$Mn^{4+}$ indicates lower covalency than in the case of $Ba_2LaNbO_6$:1%$Mn^{4+}$ ($\beta_1$=0.9691), consistent with prior assumptions (Figure 2e). All these parameter are enlisted in the Table 2.

**Table 2**. Crystal field and Racah parameters and covalency calculated for $Ba_2LaNbO_6$:$Mn^{4+}$ and $Ca_2LaNbO_6$:$Mn^{4+}$ phosphors.

| Parameter | $Ba_2LaNbO_6$:$Mn^{4+}$ | $Ca_2LaNbO_6$:$Mn^{4+}$ |
|---|---|---|
| $Dq/B$ | 3.02 | 2.73 |
| $B$ | 662 | 723 |
| $C$ | 3370 | 3068 |
| $\beta_1$ | 0.969 | 0.9472 |

A comparison of luminescence decay profiles for the $^2E$ level at 83 K shows slight deviations from exponential behavior. To quantitatively analyze the luminescence decay, average lifetimes ($\tau_{avr}$) were calculated as described in the experimental section (Eq. 1 and 2). The $\tau_{avr}$ value for $Ca_2LaNbO_6$:1%$Mn^{4+}$ is significantly longer than that for $Ba_2LaNbO_6$:1%$Mn^{4+}$, consistent with the previously reported data (Figure 2f)[54,58–61,64,68–71].

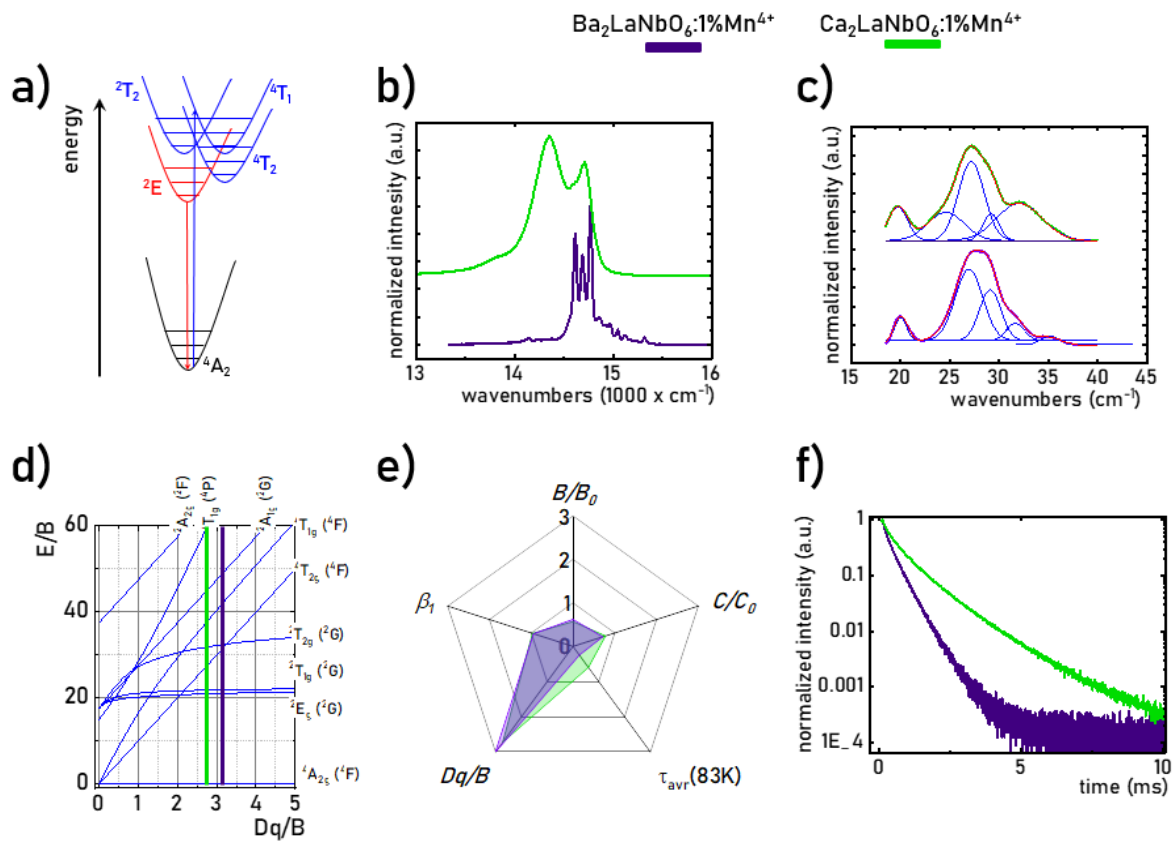

**Figure 2.** Simplified configurational-coordination diagram of $Mn^{4+}$ ions-a) the comparison of emission ($\lambda_{exc}$=359 nm for $Ba_2LaNbO_6$:1%$Mn^{4+}$ and $\lambda_{exc}$=367.5 nm for $Ca_2LaNbO_6$:1%$Mn^{4+}$)-b) and excitation ($\lambda_{em}$=680.75 nm for $Ba_2LaNbO_6$:1%$Mn^{4+}$ and $\lambda_{em}$=696.5 nm for $Ca_2LaNbO_6$:1%$Mn^{4+}$) -c) spectra of $Ba_2LaNbO_6$:1%$Mn^{4+}$ and $Ca_2LaNbO_6$:1%$Mn^{4+}$ measured at 83K; Tanabe-Sugano diagram for $3d^3$ electronic configuration-d); comparison of normalized Racah parameters $B$, $C$, $\beta_1$, $Dq/B$ and $\tau_{avr}$ (at 83K) for both phosphors -e) and the luminescence decay profiles obtained at 83K for analyzed phosphors-f).

To examine the effect of temperature on the spectroscopic properties of the analyzed materials, luminescence spectra were measured in the range of 83–403 K (Figure 3a and 3b). As evident from the presented spectra, an increase in temperature within this range does not cause any shift in the spectral position of the $^2E \rightarrow {}^4A_2$ bands for both phosphors. This indicates the absence of thermally induced distortions at the crystallographic sites occupied by $Mn^{4+}$ ions. A comparison of the integrated luminescence intensities of $Mn^{4+}$ ions in the two materials reveals a significant difference in the rate of thermal quenching (Figure 3c). For

Ba$_2$LaNbO$_6$:1%Mn$^{4+}$, an increase in temperature initially has no notable effect on the integrated emission intensity up to approximately 240 K. Above this temperature, a sharp decrease in luminescence intensity is observed. In contrast, for Ca$_2$LaNbO$_6$:1%Mn$^{4+}$, thermal quenching begins as early as 83 K, with $T_{50}$ (the temperature at which the luminescence intensity drops to 50%) determined to be 263 K, compared to $T_{50}$=322 K for Ba$_2$LaNbO$_6$:1%Mn$^{4+}$. These results suggest that Ba$_2$LaNbO$_6$:1%Mn$^{4+}$ exhibits higher thermal stability of luminescence, which is attributed to a higher activation energy ($E_a$) (Figure S5 and S6). Calculations show that $E_a$ for Ba$_2$LaNbO$_6$:1%Mn$^{4+}$ is 3620 cm$^{-1}$ (0.45 eV), whereas for Ca$_2$LaNbO$_6$:1%Mn$^{4+}$, it is only 1770 cm$^{-1}$ (0.22 eV). This finding further supports the lower crystal field strength interacting with Mn$^{4+}$ ions in Ca$_2$LaNbO$_6$:1%Mn$^{4+}$. Analysis of the luminescence decay profiles from the $^2$E level of Mn$^{4+}$ ions shows a monotonic decrease in the average lifetime ($\tau_{avr}$) with increasing temperature for both host materials (Figure 3d and 3e, Figure S7, fitting results are shown in Figures S8-S25). In Ca$_2$LaNbO$_6$:1%Mn$^{4+}$, $\tau_{avr}$ decreases from 0.6 ms at 83 K to 0.07 ms at 403 K, whereas in Ba$_2$LaNbO$_6$:1%Mn$^{4+}$, $\tau_{avr}$ reduces more drastically from 0.25 ms to 0.003 ms over the same temperature range. The rate of thermal shortening of $\tau_{avr}$ can be quantified using thermal relative sensitivity ($S_R$) as follows:

$$S_{R[\tau]} = \frac{1}{\tau_{avr}} \frac{\Delta \tau_{avr}}{\Delta T} \cdot 100\% \qquad (9)$$

where $\Delta\tau_{avr}$ corresponds to the change of $\tau_{avr}$ obtained for $\Delta T$ change in temperature. The obtained results for Ba$_2$LaNbO$_6$:1%Mn$^{4+}$ and Ca$_2$LaNbO$_6$:1%Mn$^{4+}$ reveal that in both cases $S_R$ rises up with temperature with maximal values around 403 K reaching $S_{Rmax}$=1.2% K$^{-1}$ for Ca$_2$LaNbO$_6$:1%Mn$^{4+}$ and $S_{Rmax}$=4.2% K$^{-1}$ for Ba$_2$LaNbO$_6$:1%Mn$^{4+}$ (Figure 3f).

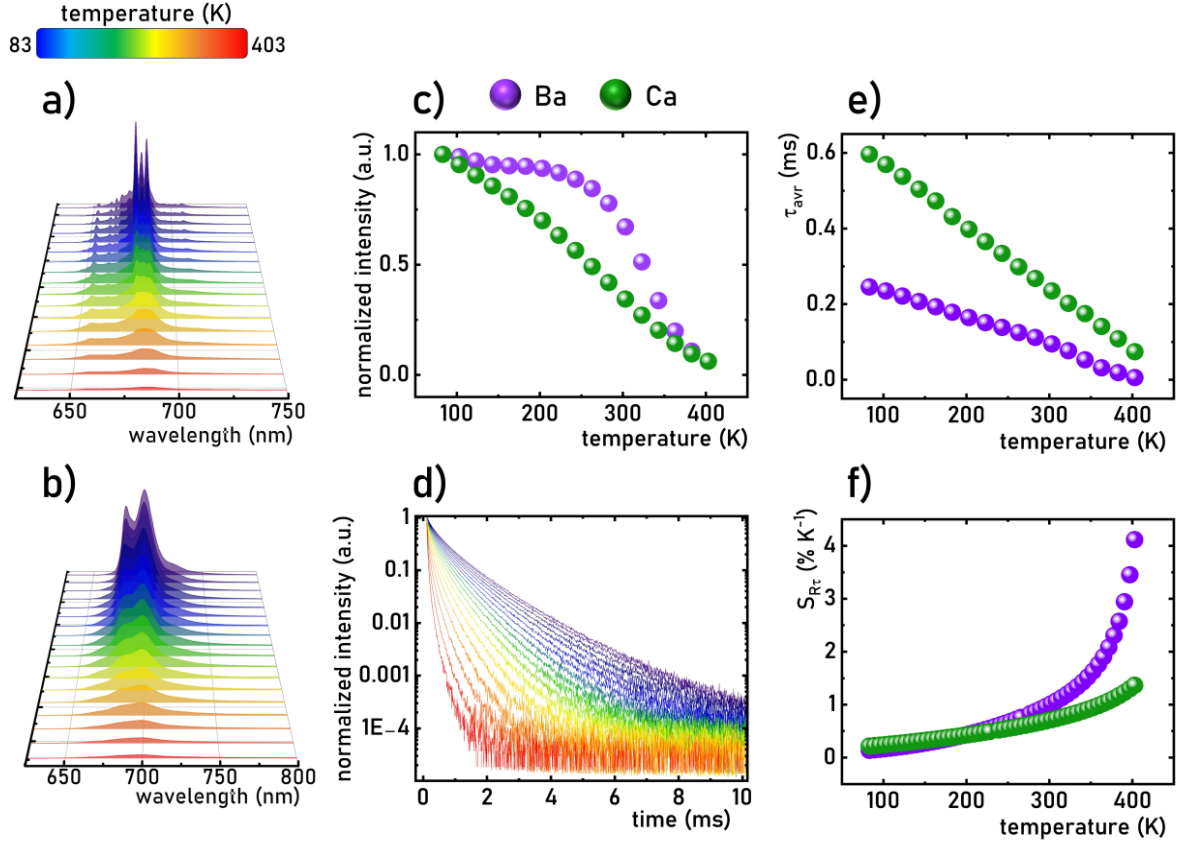

**Figure 3.** Emission spectra of $Ba_2LaNbO_6$:1%$Mn^{4+}$ -a) and $Ca_2LaNbO_6$:1%$Mn^{4+}$ -b) measured as a function of temperature ($\lambda_{exc}$=359 nm); thermal dependence of normalized (to the intensity value obtained at 83 K) emission intensities of $Ba_2LaNbO_6$:1%$Mn^{4+}$ and $Ca_2LaNbO_6$:1%$Mn^{4+}$-c); luminescence decay profiles for $^2E \rightarrow {^4A_2}$ electronic transition of $Mn^{4+}$ ions in $Ca_2LaNbO_6$:1%$Mn^{4+}$ ($\lambda_{exc}$=367.5 nm, $\lambda_{em}$= 696.5 nm)-d); and thermal dependence of average lifetimes of $^2E$ state of $Mn^{4+}$ ions in these host materials-e) and corresponding $S_R$ -f).

The thermometric approach based on $\tau_{avr}$ requires the measurement of the luminescence decay profile, which can be extremely time-consuming, particularly when two-dimensional imaging of the temperature distribution of an analyzed object is required. To address this limitation, a much simpler and faster solution-especially relevant for thermal imaging-is an approach based on the analysis of luminescence intensity recorded within two temporal gates. In this method, the thermometric parameter is defined as the ratio of luminescence intensities recorded in two time gates, and thermal distribution maps are synthesized by dividing images captured in these time windows. This approach is straightforward to implement because most cameras can adjust acquisition time, allowing gate durations to be modified without requiring complex methodologies. To verify the effectiveness of this approach, the integral luminescence intensity within different time gates of varying durations and delays relative to the excitation beam was calculated. The $^2E \rightarrow {}^4A_2$ emission intensity of $Mn^{4+}$ ions was integrated over the following temporal ranges (indicated in Figures 4a): $A$=[0-0.1] ms, $B$=[0.1-0.2] ms, $C$=[0.2-0.3] ms, $D$=[0.3-0.4] ms, $E$=[0.4-0.5] ms, $F$=[0-0.5] ms, $G$=[0.5-1.0] ms, $H$=[0-1] ms, $I$=[1-2] ms, $J$=[0-2] ms, $K$=[2-4] ms. The thermal variation of emission intensity for each gate was analyzed for $Ba_2LaNbO_6$:1%$Mn^{4+}$, as presented in Figure 4b. For shorter gates (e.g., *A, B, C, D, E* each ~0.1 ms), an initial slight decrease in intensity with increasing temperature was observed until approximately 300 K, beyond which a sharp drop in intensity occurred. This behavior is attributed to the fact that the highest luminescence intensities are observed during the initial phase of the decay profile, regardless of its overall shape. As the gate is delayed further from the excitation pulse, the initial thermal intensity variation becomes more pronounced, while the decrease in intensity beyond 300 K becomes less steep. The temperature at which this intensity reduction is observed correlates with the thermal dependence of the total luminescence intensity (Figure 3c). For gates longer than 0.5 ms, a different trend was observed. Here, the initial increase in temperature caused a sharp drop in integral intensity, with the J gate exhibiting a particularly dramatic decrease-reaching 20% of the initial intensity by 200 K. These differences in the thermal dependence of $Mn^{4+}$ emission intensities recorded at different gates highlight the feasibility of using the ratio of these intensities to remotely measure temperature. Six distinct luminescence intensity ratios (*LIR*) were proposed and defined as follows:

$$LIR_1 = \frac{A}{B} = \frac{\int_{0ms}^{0.1ms} \left[ {}^2E \rightarrow {}^4A_2 \right] dt}{\int_{0.1ms}^{0.2ms} \left[ {}^2E \rightarrow {}^4A_2 \right] dt} \qquad (10)$$

$$LIR_2 = \frac{A}{E} = \frac{\int_{0ms}^{0.1ms}\left[^2E \to {}^4A_2\right]dt}{\int_{0.4ms}^{0.5ms}\left[^2E \to {}^4A_2\right]dt} \qquad (11)$$

$$LIR_3 = \frac{A}{F} = \frac{\int_{0ms}^{0.1ms}\left[^2E \to {}^4A_2\right]dt}{\int_{0ms}^{0.5ms}\left[^2E \to {}^4A_2\right]dt} \qquad (12)$$

$$LIR_4 = \frac{F}{G} = \frac{\int_{0ms}^{0.5ms}\left[^2E \to {}^4A_2\right]dt}{\int_{0.5ms}^{1.0ms}\left[^2E \to {}^4A_2\right]dt} \qquad (13)$$

$$LIR_5 = \frac{H}{I} = \frac{\int_{0ms}^{1ms}\left[^2E \to {}^4A_2\right]dt}{\int_{1ms}^{2ms}\left[^2E \to {}^4A_2\right]dt} \qquad (14)$$

$$LIR_6 = \frac{J}{K} = \frac{\int_{0ms}^{2ms}\left[^2E \to {}^4A_2\right]dt}{\int_{2ms}^{4ms}\left[^2E \to {}^4A_2\right]dt} \qquad (15)$$

The design of the *LIRs* ensured an increase in their values across the entire analyzed temperature range (Figure 4c). For *LIR$_1$* to *LIR$_4$*, temperature increases up to 300 K had minimal impact on their values; however, beyond 300 K, a sharp rise-reaching nearly three orders of magnitude for *LIR$_4$*-was observed. Conversely, LIR$_6$ displayed a continuous increase throughout the analyzed range, with saturation occurring above 310 K. Quantitative analysis of these thermal variations was performed by calculating the relative sensitivity using a method analogous to Equation 8:

$$S_R = \frac{1}{LIR}\frac{\Delta LIR}{\Delta T}\cdot 100\% \qquad (16)$$

The thermal variability of $LIR_i$ was reflected in $S_R$ values, which peaked above 300 K (Figure 4d). The highest sensitivity, $S_{Rmax}$=17.1% K$^{-1}$, was recorded for $LIR_2$ at 330 K, more than double the sensitivity achieved with the classical $\tau_{avr}$-based approach. Importantly, by selecting appropriate time gates and thus specific $LIRs$, the temperature corresponding to maximum sensitivity can be adjusted. For example, $S_{Rmax}$ values of 10.9 % K$^{-1}$, 1.4 % K$^{-1}$, 9.5 % K$^{-1}$ and 3.2 % K$^{-1}$ were obtained for $LIR_1$, $LIR_3$, $LIR_4$ and $LIR_5$ respectively. However, the thermal operating range for these parameters remained relatively narrow (~60 K). Using $LIR_6$, however, $S_R$>1% K$^{-1}$ was achieved across a broader range of 120–270 K. This flexibility is particularly important, as it allows the thermometric performance of the luminescence thermometer to be fine-tuned by adjusting the time gates to meet specific application requirements. Notably, this effect is not unique to $Ba_2LaNbO_6$:1%$Mn^{4+}$; similar behavior and thermal variations in $LIR_i$ were observed for $Ca_2LaNbO_6$:1%$Mn^{4+}$, with maximum sensitivities of 5.2 % K$^{-1}$, 7.2 % K$^{-1}$, 0.9 % K$^{-1}$, 9.4 % K$^{-1}$ and 5.9 % K$^{-1}$ were obtained for $LIR_1$, $LIR_2$, $LIR_3$, $LIR_4$ and $LIR_5$ respectively (Figure 4e-g). It is very important to notice that the duration of the time gate affect the signal to noise ratio (SNR) (Figure S26), which affect the precision of the temperature readout. An increase of the duration of the time gates lead to an enhancement of SNR, reducing the influence of noise on temperature readouts. Slightly higher emission intensity of $Ba_2LaNbO_6$:1%$Mn^{4+}$ in respect to $Ca_2LaNbO_6$:1%$Mn^{4+}$ results in a higher observed SNR values. The results presented in this study underscore the significant potential of this approach for remote temperature sensing. The advantages of high sensitivity, ease of implementation, and compatibility with thermal imaging warrant further exploration of this methodology. While the current study focuses on microcrystalline powders doped with $Mn^{4+}$ ions, the discussed technique is equally applicable to nanocrystals. However, in such cases, additional factors must be considered, including energy diffusion processes, scattering phenomena, effective refractive index variations, and other photonic effects that can influence the luminescence kinetics of nanocrystals[79–82]. Moreover, in systems co-doped with multiple optical ions, the population pathways of the emitting levels-particularly the lifetimes of both the emitting and optically pumped levels-can alter the luminescence behavior[79–81]. Considering that the duration of the excitation pulse can also impact the observed luminescence kinetics in these systems, it is often more advantageous to utilize phosphors doped with a single type of ion for such applications.

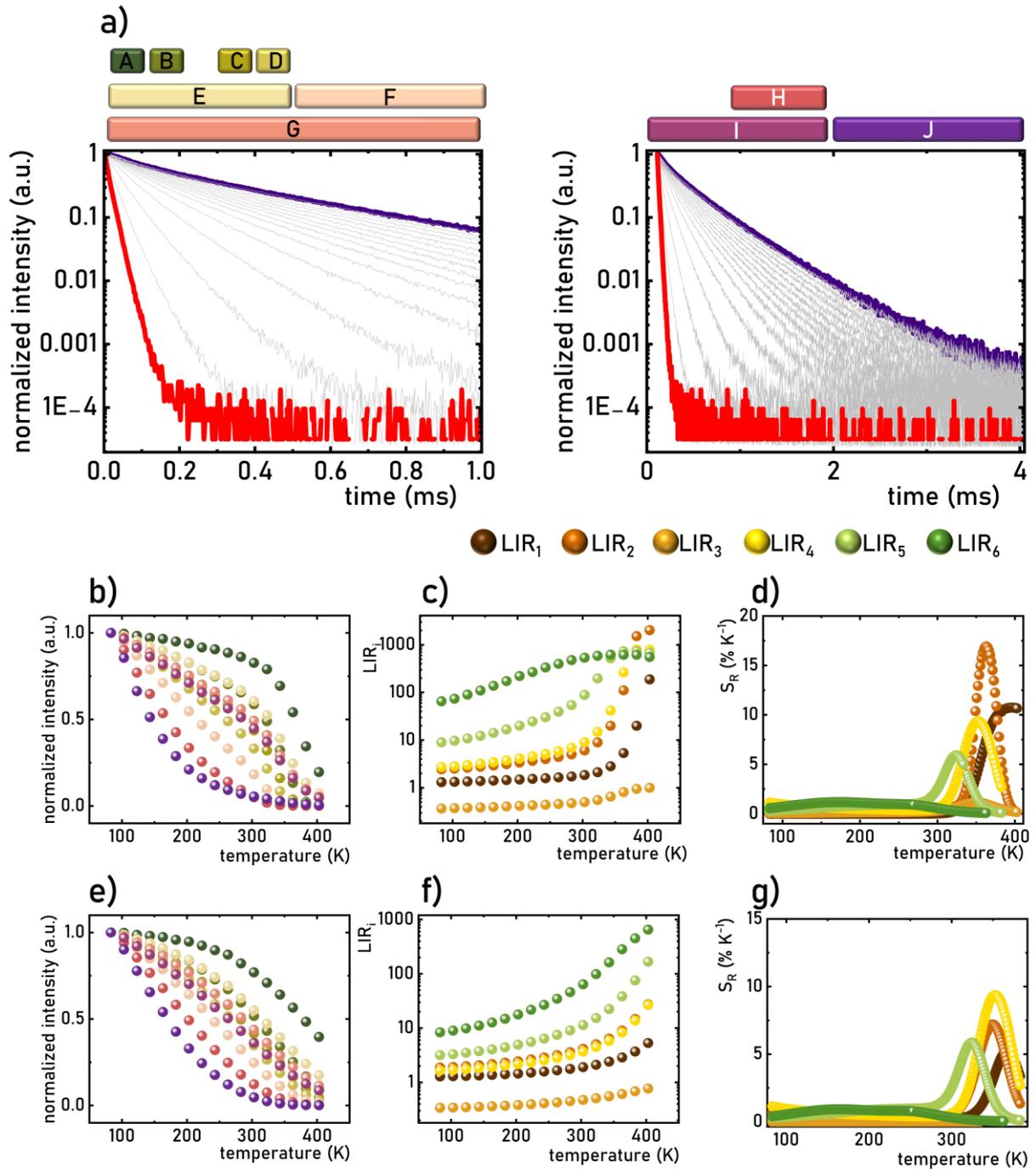

**Figure 4.** Representative luminescence decay profile of $Mn^{4+}$ ions in $Ba_2LaNbO_6$:1%$Mn^{4+}$ with marked temporal gates used in the analysis -a); the thermal dependence of emission intensity for $Ba_2LaNbO_6$:1%$Mn^{4+}$ calculated in different temporal gates – b); $LIR_i$-c) and corresponding $S_R$-d); the thermal dependence of emission intensity for $Ca_2LaNbO_6$:1%$Mn^{4+}$ calculated in different temporal gates – e); $LIR_i$-f) and corresponding $S_R$-g);

**Conclusions**

In conclusion, this paper introduces a novel methodology for determining temperature in lifetime-based luminescence thermometry by utilizing the ratio of luminescence intensities recorded at two distinct time gates. The effect of gate duration on the thermometric performance of the luminescence thermometers was investigated using two representative phosphors doped with $Mn^{4+}$ ions, differing in crystal field strength: $Ba_2LaNbO_6$:1%$Mn^{4+}$ and $Ca_2LaNbO_6$:1%$Mn^{4+}$. The study revealed that differences in the activation energy for thermal depopulation of the emitting $^2E$ state influenced the rate of thermal quenching of the $^2E \rightarrow {}^4A_2$ emission and the associated luminescence kinetics. Specifically, the higher crystal field strength ($Dq$) in $Ba_2LaNbO_6$:1%$Mn^{4+}$ resulted in stable luminescence intensity up to approximately 240 K, beyond which thermal quenching became significant. In contrast, $Ca_2LaNbO_6$:1%$Mn^{4+}$ exhibited a monotonic decrease in luminescence intensity across the entire temperature range analyzed. Thermal quenching of the $^2E$ state lifetime was observed, with values decreasing from 0.25 ms to 0.003 ms for $Ba_2LaNbO_6$:1%$Mn^{4+}$ and from 0.6 ms to 0.07 ms for $Ca_2LaNbO_6$:1%$Mn^{4+}$ within the 83–403 K range. The resulting relative sensitivity ($S_R$) of luminescence thermometers based on $\tau_{avr}$ was determined to be 4.2 % $K^{-1}$ and 1.2 % $K^{-1}$ for $Ba_2LaNbO_6$:1%$Mn^{4+}$ and $Ca_2LaNbO_6$:1%$Mn^{4+}$, respectively. However, as discussed, the classical approach-relying on measuring the luminescence decay profile-hinders thermal imaging due to the need for time-consuming point-by-point measurements. The proposed method, based on the ratio of luminescence intensities, offers a significant advantage by enabling rapid thermal imaging. Furthermore, this approach achieves considerably higher relative sensitivity $S_{Rmax}$, reaching 17.1 % $K^{-1}$ and 9.4 % $K^{-1}$ for $Ba_2LaNbO_6$:1%$Mn^{4+}$ and $Ca_2LaNbO_6$:1%$Mn^{4+}$, respectively. Importantly, adjusting the duration of the time gates allows modulation of the thermometric performance. Short time gates corresponding to the initial luminescence decay phase yield high relative sensitivities but with a limited thermal operating range due to the pronounced intensity variations at this stage. Conversely, longer time gates (1-2 ms) provide significantly lower relative sensitivities but offer a much broader thermal operating range. The advantages of the proposed method, including high relative sensitivity and suitability for fast thermal imaging, are expected to contribute to the widespread popularity of this approach in luminescence thermometry.

**Acknowledgements**

This work was supported by the National Science Center (NCN) Poland under project no. DEC-UMO-2023/49/B/ST5/03384. M.G.B. appreciates support from the Ministry of Science, Technological Development, and Innovation of the Republic of Serbia (451-03-47/2023-01/200017), the China-Serbia Intergovernmental Science and Technology Cooperation Program (Grant No. 2024[7]/6-10), the Specialized Funding Program for the Gathering of 100 Elite Talents in Chongqing, the Overseas Talents Plan (Grant No. 2022[60]) both offered by Chongqing Association for Science and Technology and the Estonian Research Council grant (PRG 2031).